# Compact radio emission indicates a structured jet was produced by a binary neutron star merger

[1,2,12]*G. Ghirlanda, [1,2,12]*O. S. Salafia, [3]Z. Paragi, [4]M. Giroletti, [5,27]J. Yang, [3]B. Marcote, [3]J. Blanchard, [6]I. Agudo, [22]T. An ,[7]†M. G. Bernardini, [8]R. Beswick, [9,26]M. Branchesi, [1]S. Campana, [10]C. Casadio, [11]E. Chassande–Mottin, [2,12]M. Colpi, [1]S. Covino, [1]P. D'Avanzo, [13]V. D'Elia, [14]S. Frey, [15]M. Gawronski, [1]G. Ghisellini, [3,16]L.I. Gurvits, [17,18]P.G. Jonker, [3,19]H. J. van Langevelde, [1]A. Melandri, [8]J. Moldon, [1]L. Nava, [12]‡A. Perego, [6,28]M. A. Perez-Torres, [20]C. Reynolds, [21]R. Salvaterra, [1]G. Tagliaferri, [4]T. Venturi, [23]S. D. Vergani, [24,25]M. Zhang

[1]Istituto Nazionale di Astrofisica - Osservatorio Astronomico di Brera, Via E. Bianchi 46, I-23807 Merate, Italy
[2]Dipartimento di Fisica G. Occhialini, Università di Milano-Bicocca, Piazza della Scienza 3, IT-20126 Milano, Italy
[3]Joint Institute for Very Long Baseline Inerferometry European Research Infrastructure Consortium (VLBI ERIC), Oude Hoogeveensedijk 4, 7991 PD Dwingeloo, The Netherlands
[4]Istituto Nazionale di Astrofisica - Istituto di radioastronomia, via Gobetti 101, I40129, Bologna, Italia
[5]Chalmers University of Technology, Onsala Space Observatory, SE-439 92, Sweden
[6]Instituto de Astrofísica de Andalucía-Consejo Superior de Investigaciones Científicas (CSIC), Glorieta de la Astronomía s/n, E-18008, Granada, Spain
[7]Laboratoire Univers et Particules de Montpellier, Universitè de Montpellier, Centre National de la Recherche Scientifique/Institute National de Physique Nucleaire et Physique des Particules (CNRS/IN2P3), place Eugéne Bataillon, F--34085 Montpellier, France
[8]Electronic Multi-Element Radio Linked Interferometer Network/Very Long Baseline Interferometry (e-MERLIN/VLBI) National facility, Jodrell Bank Centre for Astrophysics, The School of Physics and Astronomy, The University of Manchester, United Kingdom
[9]Gran Sasso Science Institute, Viale F. Crispi 7, I-67100, L'Aquila, Italy
[10]Max Planck Institute fur Radioastronomie Auf dem Huegel 69, Bonn D-53121, Germany
[11]AstroParticule et Cosmologie (APC), Université Paris Diderot, CNRS/IN2P3, Commissariat à l'énergie atomique et aux énergies alternatives/ Institute for Research on the fundamental laws of the universe (CEA/Irfu), Observatoire de Paris, Sorbonne Paris Cité, F-75205 Paris Cedex 13, France
[12]Istituto Nazionale Fisica Nucleare, Sezione di Milano Bicocca, Piazza della Scienza 3, 20126 Milano, Italy
[13]Space Science Data Center, Agenzia Spaziale Italiana (ASI), Via del Politecnico, 00133, Roma, Italy
[14]Konkoly Observatory, Magyar Tudományos Akadémia (MTA) Research Centre for Astronomy and Earth Sciences, Konkoly Thege Miklós út 15-17, H-1121 Budapest, Hungary
[15]Centre for Astronomy, Faculty of Physics, Astronomy and Informatics,Nicolaus Copernicus University, Grudziadzka 5, 87-100 Torun, Poland
[16]Department of Astrodynamics and Space Missions, Delft University of Technology, Kluyverweg 1, 2629 HS Delft, The Netherlands
[17]Space Research Organisation of the Netherlands (SRON), Netherlands Institute for Space Research, Sorbonnelaan 2, 3584 CA Utrecht, The Netherlands

[18]Department of Astrophysics, Institute for Mathematics, Astrophysics and Particle Physics (IMAPP), Radboud University, P.O. Box 9010, 6500 GL Nijmegen, the Netherlands
[19]Sterrewacht Leiden, Leiden University, P.O. Box 9513, NL-2300 RA Leiden, The Netherlands
[20]Commonwealth Scientific and Industrial Research Organization (CSIRO) Astronomy and Space Science, PO Box 1130, Bentley WA 6102, Australia
[21]Istituto Nazionale di Astrofisica, Istituto di Astrofisica Spaziale e Fisica cosmica (IASF), via E. Bassini 15, 20133 Milano, Italy
[22] Shanghai Astronomical Observatory, Key Laboratory of Radio Astronomy, Chinese Academy of Sciences, 200030 Shanghai, China
[23]Galaxies, Etoiles, Physique et Instrumentation (GEPI) Observatoire de Paris, CNRS UMR 8111, Meudon, France
[24]Xinjiang Astronomical Observatory, Chinese Academy of Sciences, 150 Science 1-Street, Urumqi 831001, China
[25]Key Laboratory for Radio Astronomy, Chinese Academy of Sciences, 2 West Beijing Road, Nanjing 210008, China
[26]INFN - Laboratori Nazionali del Gran Sasso, I-67100 L'Aquila, Italy
[27]Yunnan Observatories, Chinese Academy of Sciences, 650216 Kunming, Yunnan, China
[28]Departamento de Física Teórica, Facultad de Ciencias, Universidad de Zaragoza, E-50019, Spain.

∗To whom correspondence should be addressed; E-mail:
giancarlo.ghirlanda@brera.inaf.it, omsharan.salafia@brera.inaf.it

† Present address: Istituto Nazionale di Astrofisica (INAF) – Osservatorio Astronomico di Brera, Italy

‡ Present address: Physics Department, University of Trento, via Sommarive 14, I-38123 Trento, Italy

**The binary neutron star merger event GW170817 was detected through both electromagnetic radiation and gravitational waves. Its afterglow emission may have been produced by either a narrow relativistic jet or an isotropic outflow. High spatial resolution measurements of the source size and displacement can discriminate between these scenarios. We present Very Long Baseline Interferometry observations, performed 207.4 days after the merger, using a global network of 32 radio telescopes. The apparent source size is constrained to be smaller than 2.5 milliarcseconds at the 90% confidence level. This excludes the isotropic outflow scenario, which would have produced a larger apparent size, indicating that GW170817 produced a structured relativistic jet. Our rate calculations show that at least 10% of neutron star mergers produce such a jet.**

**One Sentence Summary:** Size measurement through worldwide radio telescope array proves that a relativistic jet successfully emerged from the neutron star merger GW170817

The binary neutron star merger GW170817 was detected in both gravitational waves (GW) (*1*) and electromagnetic (EM) emission (*2*). Less than two seconds after the detection of the GW signal, a weak short duration γ-ray burst (GRB 170817A) was observed (*3, 4*). Eleven hours later, electromagnetic observations from ultraviolet to near-infrared wavelengths (*2*) pinpointed the host galaxy as NGC 4993, at ~ 41 megaparsecs (Mpc) distance. The temporal and spectral properties of this emission component reflect those expected for a kilonova, the radioactive-decay-powered emission from material ejected during and after a neutron star merger (*5, 6*). 9 and 16 days after the GW event, X-ray (*7, 8*) and radio (*9*) emissions were detected. These are interpreted as the afterglow of GRB 170817A. Monitoring of the afterglow with radio, optical and X-ray telescopes showed a slow achromatic increase in flux (*10*) (F ∝ $t^{0.8}$, where F indicates the flux and t the time elapsed since GW170817) until ~ 150 days post-merger (*11–13*). After this epoch, the flux began to decrease (*14, 15*).

Interpretation of the long-lived radio, optical and X-ray emission, has invoked the launch of a jet from the remnant of the merger. The jet drills into the surrounding kilonova material ejected shortly beforehand. Either the jet successfully breaks through the ejecta, developing an angular structure [i.e. the energy and velocity scale with the angular distance $\theta$ from the jet axis (*27*)], or it fails to break out, depositing all its energy into the ejecta and forming a hot cocoon which subsequently expands due to its high pressure (*16–19*). In the latter case the energy is expected to be distributed over a wide opening angle, and the expansion velocity is expected to be lower with respect to the jet scenario. Owing to the angular structure, the successful jet scenario is often called a structured jet (*20, 21*) while the unsuccessful jet scenario is sometimes referred to as a choked jet or cocoon.

The X-ray, optical and radio brightness as function of time (i.e. light curve) of GRB 170817A (Figure 3), up to ~230 days (*15, 22*) does not distinguish the two scenarios: with reasonable parameters, both models are consistent with those observations. Independent constraints on the geometry of the relativistic outflow can be obtained through polarization measurements and/or interferometric imaging (*23–26*). Due to the higher velocity and narrower opening angle, a structured jet is expected to have a larger displacement from the merger location and at ~200 days was predicted to be compact, with an angular size smaller than 2 milliarcseconds (mas) (*23, 25*). Conversely, a choked jet cocoon would have a smaller displacement (no detectable displacement and a ring image for a perfectly isotropic outflow) and a larger (>2 mas) apparent angular size (Figure 2). A recent measurement of the displacement of the source apparent position, by 2.67 ± 0.3 mas in 155 days (*22*), strongly supports the structured jet scenario. However, those data do not have sufficient resolution to determine the apparent size. We have used global Very Long Baseline Interferometry (VLBI) observations to place tighter limits on the source angular size, providing an independent constraint on the source structure to that obtained from the apparent motion alone (*22*).

We observed GRB 170817A on 2018 March 12-13, 207.4 days after the GW/GRB detection, using 32 radio telescopes spread over five continents. The longest baseline producing useful data was 11878 km between Hartebeesthoek (South Africa) and Fort Davis (USA). Observations were performed at a central frequency of 4.85 GHz (wavelength of 6.19 cm) with a total bandwidth of 256 MHz. The total on-source time was 7.8 hours (*27*).

The observations show a source at the sky position Right Ascension (RA) = $13^h09^m48^s.06880 ± 0^s.00002$, Declination (Dec) = $−23°22'53''.390765 ± 0.00025''$ [J2000 equinox, 1σ statistical uncertainty (*27*)]. This is within the uncertainty on the position of the optical source (*28*) and compatible with the radio position of the source obtained with the High Sensitivity Array (HSA) (*22*). With respect to the HSA observation at 75 days after the GW event (*22*), our position, measured at 207.4 days, is displaced by δRA(207.4 d − 75 d) = 2.44 ± 0.32 mas and δDec(207.4 d − 75 d) = 0.14 ± 0.47 mas. With respect to the HSA observation at 230 days (*22*), we measure δRA(230 d − 207.4 d) = 0.46 ± 0.34 mas and δDec(230 d − 207.4 d) = 0.07 ± 0.47 mas (1σ statistical uncertainties). Our global VLBI observation, performed shortly after the source flux density peak (Figure 3), has a position intermediate between those two HSA observations (Figure 1A) and matches the apparent superluminal motion seen in the HSA data (*22*). We measure a peak brightness of 42 ± 8 microJanskys (μJy) beam$^{-1}$ at 5 GHz. This is consistent with the value 47 ± 9 μJy obtained by interpolating the closest previously-published radio observations (*11, 15*). We also obtained quasi-simultaneous observations with the electronic Multi-Element Radio Linked Interferometer Network (e-MERLIN) array (*27*), measuring a consistent peak brightness upper limit of <60 μJy beam$^{-1}$ (at 3σ significance).

The effective angular resolution of the global VLBI reconstructed image is 1.5 × 3.5 mas (*27*). The source in the image appears compact and apparently unresolved with such resolution (Figure 1A). We calculate (*27*) that the source size at 207 days as measured from the global VLBI image is smaller than 2.5 mas at the 90% confidence level (Figure S4).

We compared our data with four possible models of the outflow (*27*), consisting of a successful jet and three variants of the choked jet scenario (Figure 2). For the successful jet, model parameters are determined by simultaneous fitting (similar to (*29*)) of the 3 GHz, optical and X-ray light curves and of the observed centroid displacement, obtained by comparing the position in our observations to those in the HSA observations (*22*). Adopting jet parameters inferred indirectly from previous observations (*12, 13, 18*) yields similar (within 20%) image sizes, thus our conclusions are not sensitive to the particular parameters chosen. The three choked jet models are characterized by different degrees of anisotropy, parametrized by the outflow collimation angle $θ_c$ ranging from 30 to 60 degrees, all with a viewing angle of 30 degrees. All three choked jet models match the observed multi-wavelength light curves. However, their image sizes differ in the three cases, and are all larger than the successful jet.

All three choked jet cocoon models are excluded by the image size measured in our observations (*27*). We therefore favor the structured jet model for GRB170817A: a successful jet with a structured angular velocity and energy profile, featuring a narrow ($θ_c$ = 3.4 ± 1°) and energetic ($E_{iso,core}$ = $E_c$ = $2.5^{+7.5}_{-2.0}×10^{52}$ erg, where $E_{iso,core}$ is isotropic equivalent energy of the jet core) core seen from a viewing angle of ~15 degrees (see *27* for a discussion on the uncertainty on the viewing angle). The synthetic image for this model (Figure 1B) is similar to the observed image (Figure 1A). The energy and bulk velocity of the jet material decrease steeply away from the jet axis in our model (*27*), producing a sheath of slower material surrounding the core. The isotropic equivalent luminosity $L_{iso}$~$10^{47}$ erg s$^{−1}$ of GRB 170817A (*3, 4*) is lower than a typical short GRB. This gamma-ray emission was probably not produced by the jet core, because its emission would have been too narrowly beamed (due to relativistic effects) to intercept our line of sight. Instead, we infer that the gamma rays were emitted from the part of the sheath moving in our direction; in this case the slowly rising multi-wavelength emission (*10, 11, 13*) was due to the subsequent deceleration of parts of the sheath located progressively closer to the jet core. The flattening (*13*) and subsequent peak (*15*) of the light curve (Figure 3) then mark the time when the emission becomes dominated by the jet core.

If such a jet were observed on-axis, its gamma-ray emission would have had an isotropic equivalent luminosity $\geq 10^{51}$ erg s$^{-1}$ (assuming 10% efficiency in the conversion of kinetic energy to radiation). Studies of the short GRB (sGRB) luminosity function (*29, 31*) indicate that the local rate of sGRBs with $L_{iso} > 10^{51}$ erg s$^{-1}$ is ~ 0.5 yr$^{-1}$ Gpc$^{-3}$. Assuming that all sGRB jets have a similar (i.e. quasi-universal (*32*)) structure, and that sGRBs with $L_{iso} > 10^{51}$ erg s$^{-1}$ are produced by jets whose core points towards Earth, the rate of lower luminosity events depends on the jet structure (*33*) due to the larger number of events visible from larger viewing angles $\theta_v$. For a structured jet whose luminosity scales as a power-law (*32*) $L(\theta_v) \propto (\theta_v/\theta_c)^{-\alpha}$, as a function of the angular distance $\theta$ from the jet axis, the local rate $R_0(>L)$ of events with luminosity larger than $L$ is shown in Figure 4 for $\alpha$ = 2, 3, 4. The rate of GRBs with luminosity as low as GRB 170817A (*34*) (Figure 4) is consistent with the expected luminosity function of structured jets. Comparing the resulting rate of jets to the local rate of binary neutron star (NS–NS) mergers, $R_{NS-NS} = 1540^{+3200}_{-1220}$ yr$^{-1}$ Gpc$^{-3}$ as estimated from GW data (*1*), we argue that at least 10% of NS-NS mergers launch a jet which successfully breaks out of the merger ejecta.

**Acknowledgments:** We thank M. E. Ravasio, I. Andreoni and A. Deller for help in cross-checking the AGN position. We are grateful to M. Orienti for helpful discussion about the data analysis. The European VLBI Network is a joint facility of independent European, African, Asian, and North American radio astronomy institutes. Scientific results from data presented in this publication are derived from the EVN project codes GG084 (PI G. Ghirlanda), RG009 (PI G. Ghirlanda), EP105 (PI Z. Paragi) and from the e-MERLIN project code CY6213 (PI G. Ghirlanda).

**Funding:** The National Institute of Astrophysics is is acknowledged for PRIN-grant (2017) 1.05.01.88.06. The Italian Ministry for University and Research (MIUR) is acknowledged through the project "FIGARO" (Prin-MIUR) grant 1.05.06.13. ASI is acknowledged for grant I/004/11/3. The research leading to these results has received funding from the European Commission Horizon 2020 Research and Innovation Programme under grant agreement No. 730562 (RadioNet). The Spanish Ministerio de Economa y Competitividad (MINECO) is acknowledged for financial support under grants AYA2016-76012-C3-1-P, FPA2015-69210-C6-2-R, and MDM-2014-0369 of ICCUB (Unidad de Excelencia "Mara de Maeztu"). M.A.P.-T. acknowledges support from the Spanish MINECO through grants AYA2012-38491-C02-02 and AYA2015-63939-C2-1-P. TA is supported by the National Key R&D Programme of China (2018YFA0404603). ECM acknowledges support from the European Union's Horizon 2020 research and innovation program under grant agreement No.653477. SF thanks the Hungarian National Research, Development and Innovation Office (OTKA NN110333) for support. The Long Baseline Array is part of the Australia Telescope National Facility which is funded by the Australian Government for operation as a National Facility managed by CSIRO. e-MERLIN is a National Facility operated by the University of Manchester at Jodrell Bank Observatory on behalf of STFC.

**Author contributions:** All authors contributed to the text and supplementary material, the design of the science case, and the technical definition and implementation of the observations. G. Ghirlanda and O. S. Salafia coordinated the work and performed the interpretation of the results through discussions with M. G. Bernardini, S. Campana, E. Chassande-Mottin, M. Colpi, S. Covino, P. D'Avanzo, V. D'Elia, G. Ghisellini, A. Melandri, L. Nava, A. Perego, R. Salvaterra, G. Tagliaferri, S. D. Vergani. The global-VLBI data were analysed by Z. Paragi with contributions by B. Marcote and J. Blanchard and blindly cross-checked through an independent analysis performed by M. Giroletti and J. Yang. R. Beswick and J. Moldon performed the eMERLIN data reduction and analysis. M. Giroletti and M. A. Perez—Torres performed analysis of the RG009-EVN observation (first epoch), which was independently and analysed by Z. Paragi.  E. Chassande-Mottin and M. Branchesi provided expertise on the GW event. T. Venturi provided support for the observations. Numerical codes were implemented by O. S. Salafia for the production of the images and light curves of the models, and G. Ghirlanda for the sGRB rates, both with contributions and discussions with G. Ghisellini and M. Colpi. A. Melandri and P. D'Avanzo contributed multi-wavelength data for the light curve. P. G. Jonker, I. Agudo, T. An, C. Casadio, S. Frey, M. Gawronski, L. I. Gurvits, H. J. van Langevelde, C. Reynolds, M. Zhang provided comments on the text.

**Competing interests:** S. D. Vergani is also affiliated with the Istituto Nazionale di Astrofisica (INAF) – Osservatorio Astronomico di Brera, Italy.

**Data and materials availability:** EVN data are available from the archive http://archive.jive.nl/scripts/portal.php under project codes GG084, RG009 & EP105. The e-MERLIN data and scripts are available from *(37)*. Our analysis and modelling software codes, and output model images, are available from *(38)*.

**Supplementary Materials:**

Materials and Methods

Figures S1-S6

Tables S1-S2

References (*39-61*)

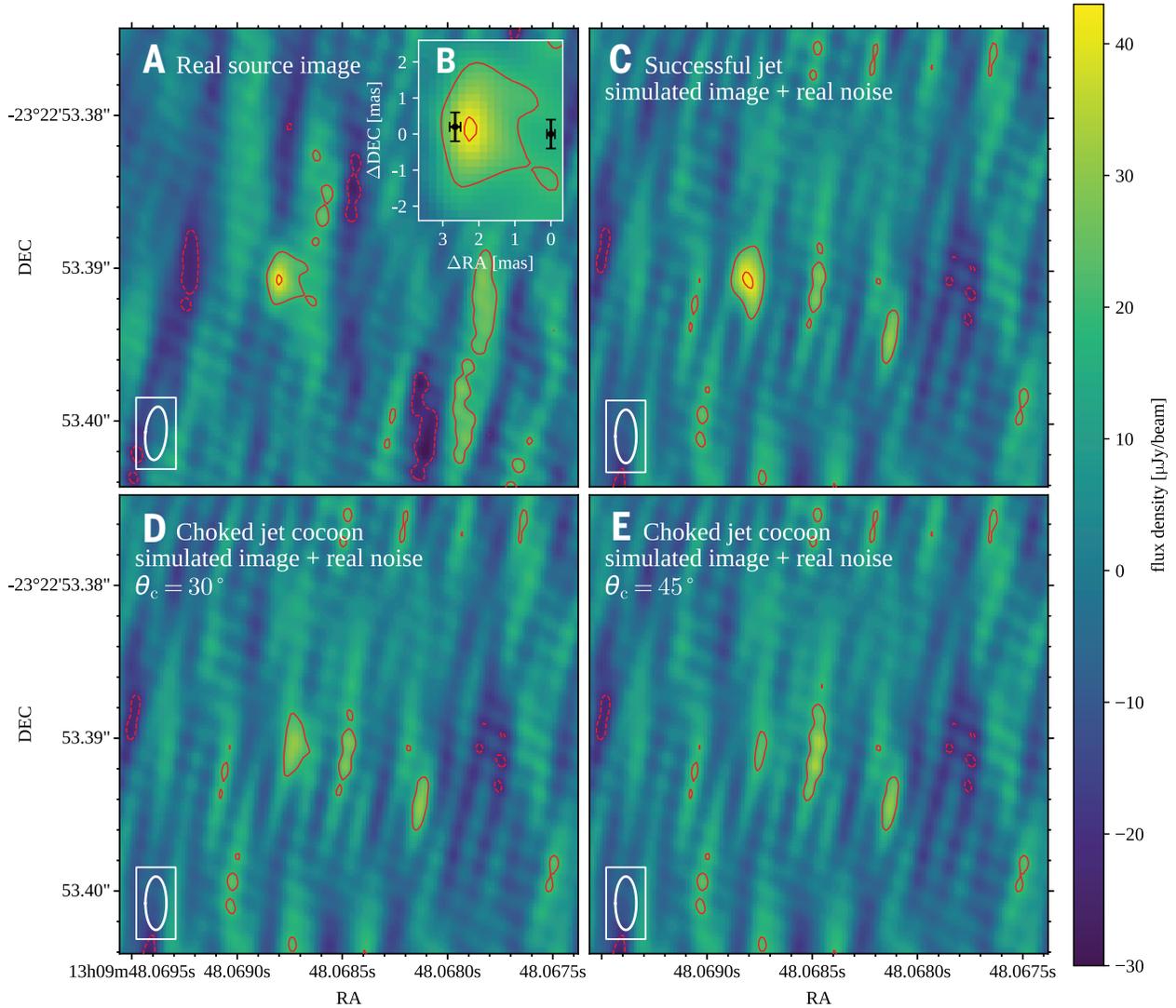

Fig. 1. Observed and simulated radio images of GRB170817A. (A) radio image from our global-VLBI observation (measured brightness rms of 8 μJy beam$^{-1}$). Red contours (dashed for negative values) indicate brightness levels of -20, 20 and 40 μJy beam$^{-1}$. The beam size (3.5×1.5 mas) is illustrated by the ellipse in the bottom left. (B): a zoom on the position of the source, with black error bars showing previously reported (22) centroid positions at 75 days and 230 days after the merger. The source is moving to the left in this orientation. Axes show the projected distance in milliarcseconds from the position at 75 days. (C) same as panel A, but showing a simulated radio image for the structured jet model, convolved to the same beam as the observation, with real noise added. (D) same as B, but for the choked jet cocoon model with $\theta_c = 30°$. (E) same as D, but for $\theta_c = 45°$. The structured jet model most closely matches the observations.

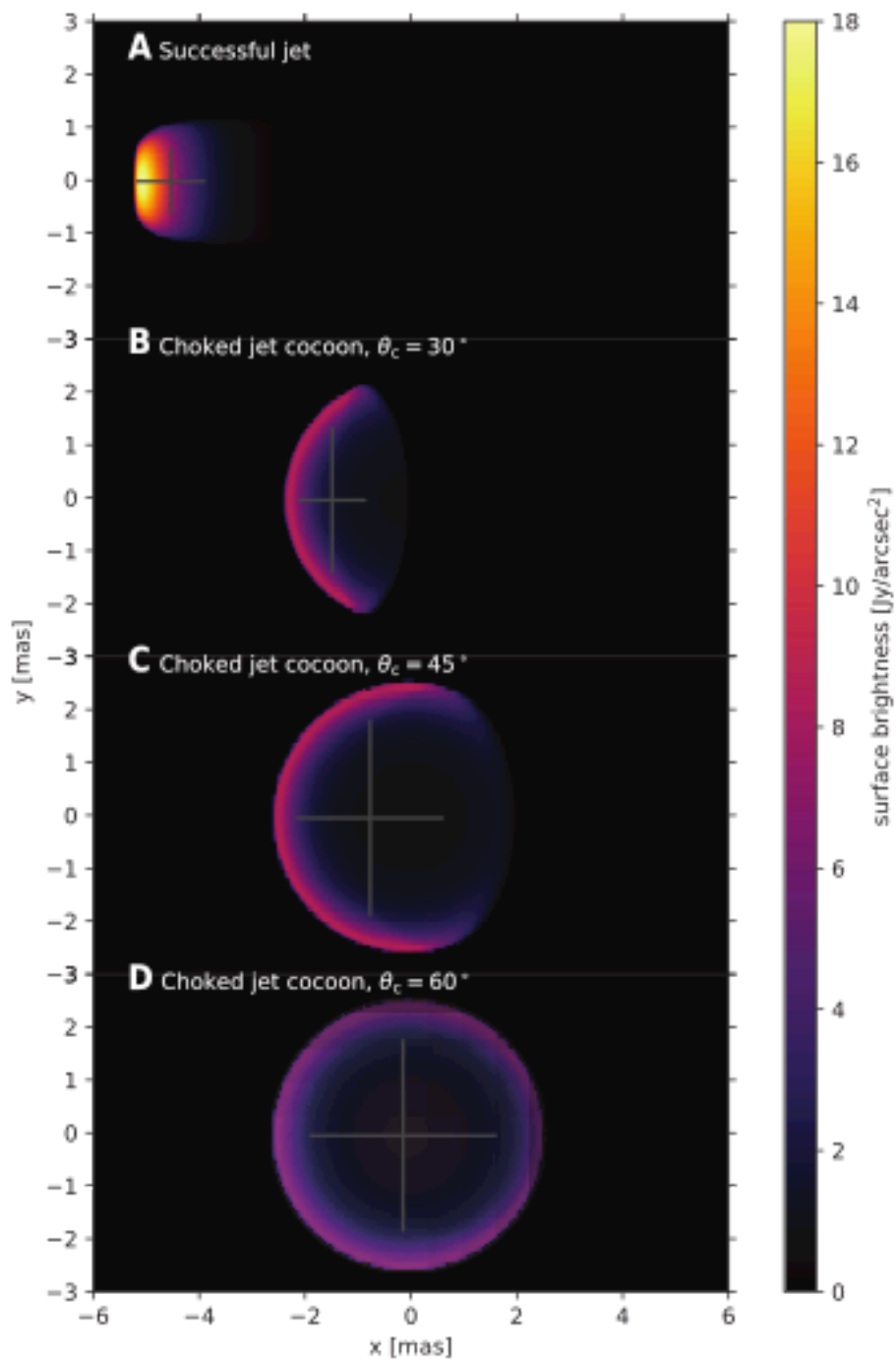

**Fig. 2. Predicted source images for our four models.** (A) predicted radio brightness distribution at 207.4 d for the structured jet model. (B-D) same as A, but for the choked jet cocoon model with effective opening angles of $\theta_c = 30°$ $(B), 45°(C), 60°(D)$. Grey crosses show the positions and sizes (full widths at half maxima) of elliptical Gaussian fitting of the images. The coordinate origin, in each image, is the projected position of the binary neutron star merger (*27*).

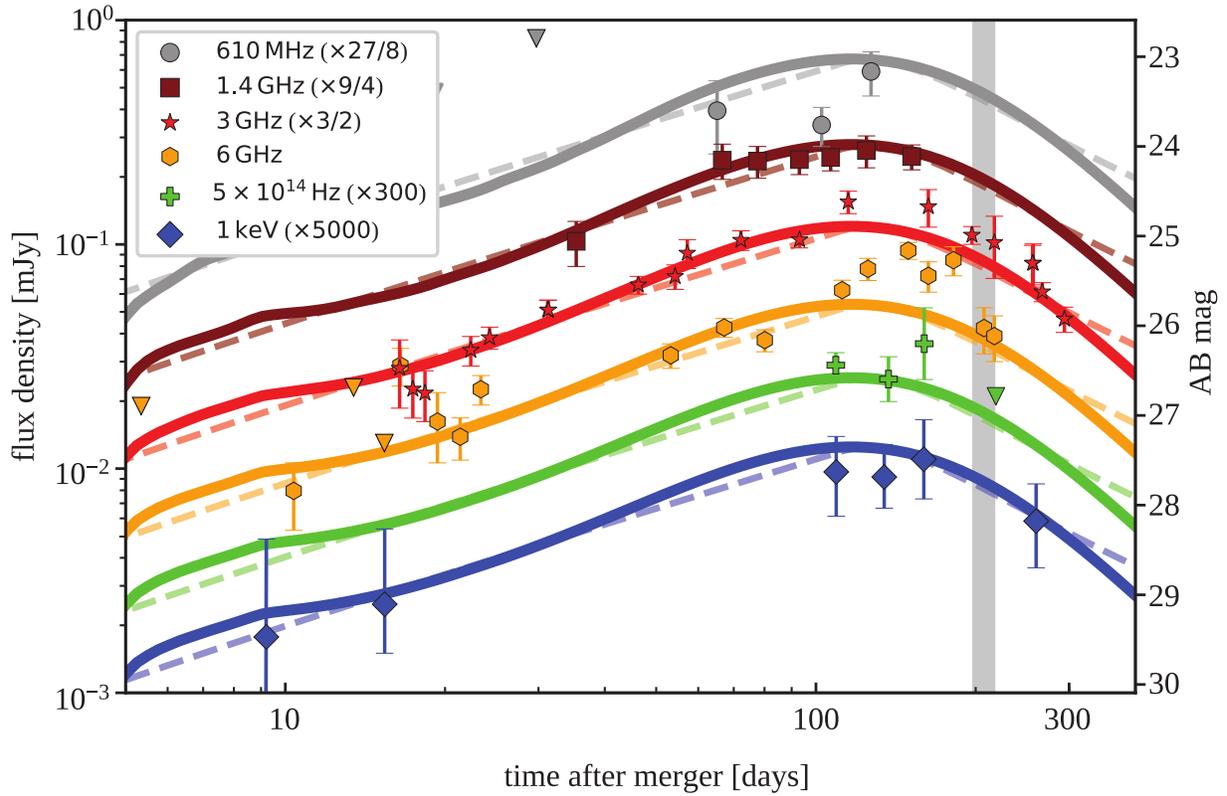

**Fig. 3. Multiwavelength light curves of GRB170817.** Model curves are shown for the structured jet model (solid lines) and the choked jet cocoon with velocity profile (dashed lines). Models are parametrised as described in (*27*). Upper limits are shown by downward triangles. Data are taken from *(13,14, 15, 35* including the optical detection of the afterglow of GRB170817A *36*). The shaded grey vertical bar marks the date of our global VLBI observation. Data and model curves are shifted by multiplicative factors (given in the legend) for ease of display.

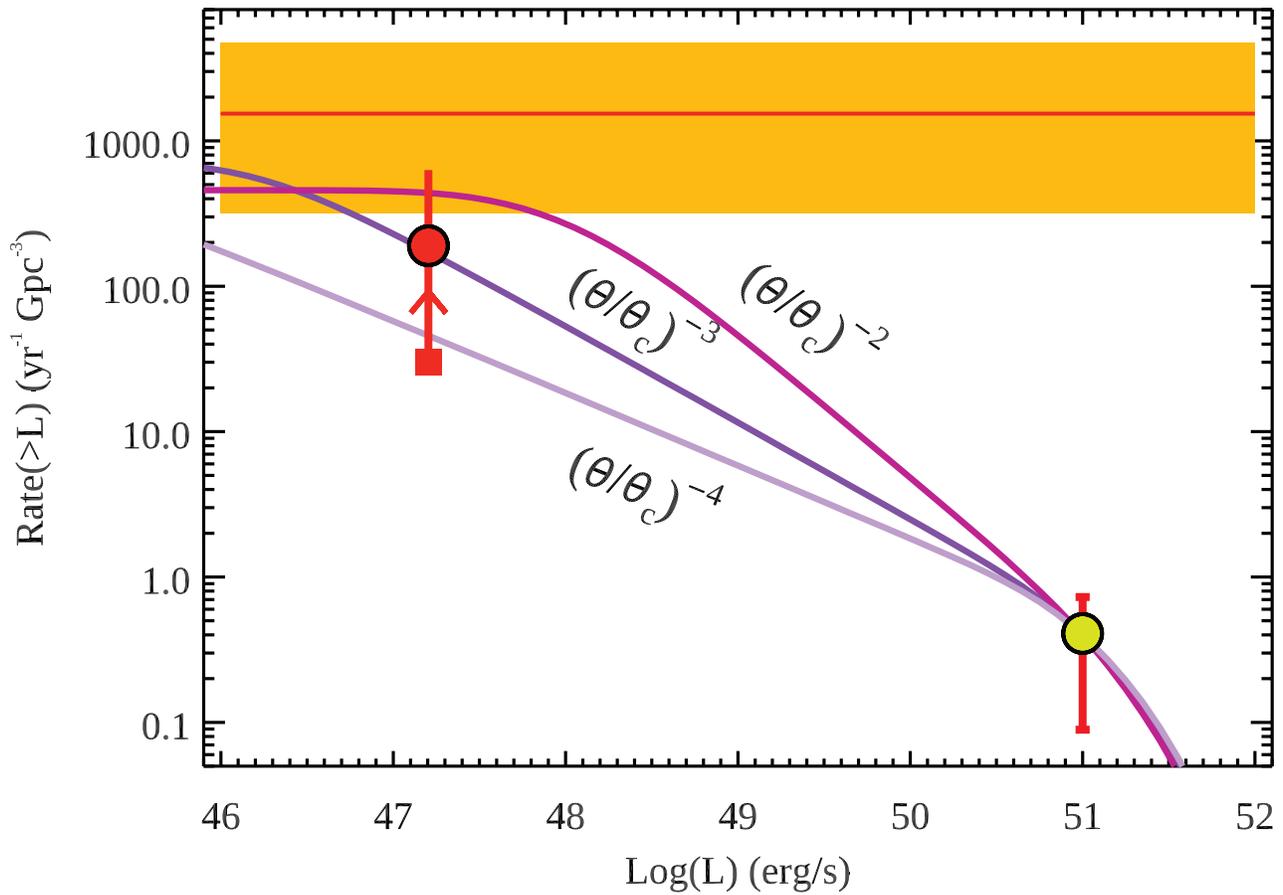

**Fig. 4. Short GRB rate as a function of luminosity.** The rate of short GRBs with isotropic equivalent luminosity $L_{iso} > 10^{51}$ erg s$^{-1}$ (yellow filled symbol - *27*) is compared with the expected rate of short GRBs similar to GRB 170817A (solid red symbol (*34*)). We consider this a lower limit: GRB170817A, detected by the Fermi spacecraft, is the only one of its class with an associated GW event, but Fermi could have detected similarly dim events without an associated GW event. Lines show predictions for different jet structures, consistent with the estimate based on the detected luminosity of GRB 170817A. The solid red horizontal line (the orange shaded region is its 1σ uncertainty) shows the rate of binary neutron stars (BNS) mergers inferred from GW data alone (*1*).

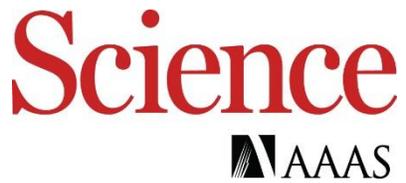

Supplementary Materials for

**Compact radio emission indicates a structured jet was produced by a binary neutron star merger**

G. Ghirlanda*, O. S. Salafia*, Z. Paragi, M. Giroletti, J. Yang, B. Marcote, J. Blanchard,
I. Agudo, T. An, M. G. Bernardini, R. Beswick, M. Branchesi, S. Campana, C. Casadio,
E. Chassande–Mottin, M. Colpi, S. Covino, P. D'Avanzo, V. D'Elia, S. Frey, M. Gawronski,
G. Ghisellini, L.I. Gurvits, P.G. Jonker, H. J. van Langevelde, A. Melandri, J. Moldon, L. Nava,
A. Perego, M. A. Perez-Torres, C. Reynolds, R. Salvaterra, G. Tagliaferri,   T. Venturi,
S. D. Vergani, M. Zhang

*Correspondence to: giancarlo.ghirlanda@brera.inaf.it, omsharan.salafia@brera.inaf.it

**This PDF file includes:**

Materials and Methods
Figs. S1-S6
Tables S1-S2

**Materials and Methods**

Global VLBI observations, data reduction and analysis.

The observations were carried out on 2018 March 12–13 at 5 GHz with a global array of radio telescopes including the Australia Telescope Compact Array (ATCA) (phased array of 5×22 m), Mopra, Ceduna and Hobart from the Long Baseline Array (LBA); Tianma, Urumqi, Badary, Kunming, Hartebeesthoek, Zelenchukskaya, Noto, Medicina, Effelsberg, Jodrell Bank MkII, Irbene 16m, Onsala 25m, Yebes, Torun, and Westerbork (single dish) from the European VLBI Network (EVN); eight telescopes from the Very Long Baseline Array (VLBA: Hancock, North Liberty, Fort Davis, Los Alamos, Kitt Peak, Owens Valley, Brewster, and Mauna Kea) and the R.C. Byrd Green Bank Telescope. The Karl G. Jansky Very Large Array (VLA) did not take part because of a power failure. Westerbork, Torun, ATCA, Mopra and Ceduna recorded at a total bit rate of 1 Gbit/s, the rest of the array at 2 Gbits$^{-1}$ recording. No fringes were detected on the baselines to Hobart, and the sensitivity on baselines to Ceduna was very low. The sensitivity for all the EVN telescopes recording at 2 Gbit/s was lower than expected for certain subbands in this observing session of 2018, due to issues with the calibration control in the digital baseband converters (DBBC). The data were correlated at the Joint Institute for VLBI European Research and Infrastructure Consortium (ERIC-JIVE) in Dwingeloo, the Netherlands, with the EVN Software Correlator (SFXC; (*39*)). The Irbene 16m data disk did not arrive in time for rapid correlation. The total recorded bandwidth was 256 MHz per polarization; most telescopes observed both left- and right-hand circular polarizations (Ceduna had right circular polarization only, Mopra operated in two linear polarizations). The bandwidth was divided into 16×16 MHz sub-bands during correlation. The correlator integration time was 1s and the spectral resolution was 0.5 MHz. The resulting field of view for an array extending to 12000 km was 8.25 arcseconds in radius (10% loss in amplitude), limited by bandwidth smearing. As a result, the NGC 4993 nucleus in the field of view of GRB170817A was somewhat smeared for the longest baselines in our data.

Due to the low declination – and thus very low elevations for telescopes in the northern hemisphere – the phase-referencing observations were quite challenging. We used a phase-referencing cycle of 1.5 minutes on calibrator and 2.5 minutes on the target (both include slewing times). The phase-reference source was J1311−2329 (NVSS J131137-232954), separated by 26 arcmin from both the target and the phase-referencing check source J1312−2350. The correlation position of J1311−2329 (RA = $13^h11^m37^s.413987$, Dec = $−23°29'56''.64651$, J2000 equinox) was determined by our earlier observations on 11 October 2017 (project EP105A), with respect to the bright International Celestial Reference Frame (ICRF) source J1303−2405 (2MASS J13031963-2405031), which was used to calibrate previous GW170817 observations (*40, 41*). While our initial position on the phase-reference calibrator was accurate to a few milliarcseconds, using the very nearby check source J1312−2350 (WISE J131248.76-235047.3) allowed us to determine precise positions in the ICRF. We used J1337−1257 (2MASS J1337397-125725) as a fringe-finder.

The data were reduced using the Astronomical Image Processing System (AIPS) package (*42*). As a first step, we followed the EVN Data Reduction Guide (*44*). We used the Interferometry Data Interchange (IDI) Flexible Image Transport System (FITS - *43*) files and the pipeline (*45*) calibration tables from the EVN data repository. We copied over the second version of the pipeline calibration table (CL.2), the flag as well as the bandpass tables (FG.1, BP.1) to the data. In CL.2 parallactic angle correction and a-priori amplitude calibration information (using the system temperature $T_{sys}$ and gains measured at the telescopes) are included. We ran TECOR to correct for the signal propagation delays caused by the ionosphere, using *ionex* files (*42*) for the days of the observations. The instrumental delay and phase offsets between the different sub-bands were removed using 4 short scans of the calibrator sources (to correct for the different parts of the array) with FRING. In the final fringe-fitting in order to determine the residual delays, rates and phases, we combined all the sub-bands to increase signal-to-noise ratio. For the GRB170817A field and for the J1312−2350 check source we interpolated solutions from the reference source J1311−2329 using CLCAL. The calibration was applied to all sources and the frequencies within the sub-bands were averaged with SPLIT. To obtain the frequency-averaged data for NGC 4993, we shifted the GRB 170817A field phase-centre to the AGN position by UVFIX before SPLIT. Imaging and model-fitting were carried out by DIFMAP (*46*); we read the resulting images back to AIPS and fitted them with JMFIT.

We detected GRB170817A, the AGN in NGC 4993, and the phase-reference check source J1312−2350. The J1312−2350 data showed that phase-referencing worked, with coherence losses not exceeding the value of 10–15% typically seen in VLBI under good atmospheric conditions. While the errors affecting the visibilities are station-dependent, the delays and delay rates scale with baseline length, and therefore the visibility phases wrap with frequency and time much more rapidly on the longest baselines. This results in phase errors (to which the longest baselines are much more prone to) that blur the image, and results in a drop in source peak brightness at a level of ~10-15% in typical observing conditions (see for example (*48*)). Our a-priori reference position of J1311−2329 was however erroneous by about 1.5 milliarcseconds, as is apparent by the comparison of our obtained positions with published ones (for GRB 170817A (*22*), for NGC 4993 (*49*), and for J1312−2350 compared to the known ICRF position). After re-referencing our GRB170817A position to the known position of J1312−2350, we obtain a source position RA = $13^h09^m48^s.06880 \pm 0^s.00002$, DEC = $-23°22'53''.390765 \pm 0''.00025$, where the quoted uncertainties are the root sum square of the statistical errors on the positions of GRB 170817A and J1312-2350 as reported by JMFIT (see Table S2 for a summary of the source positions before and after applying the shift). The source GRB 170817A is the brightest in a field of view of 1×1 arcseconds in our naturally weighted dirty map, with a peak brightness of 42 ± 8 μJy beam$^{-1}$. It remains the brightest source also after applying a Gaussian taper to the uv-data. The position is within the 1-sigma uncertainty of the optical position (*28*), and within 0.5 milliarcseconds of the radio position in April 2018 (*22*).

e-MERLIN coordinated observations: data analysis and results.

Twelve coordinated observations with the e-MERLIN array were made between 8 and 22 March 2018 (project CY6213) in support of the global VLBI observations. All data were correlated in real-time with a total bandwidth of 512 MHz covering the range 5 to 5.5 GHz, divided into four spectral windows each with 512 channels per polarisation. The observing band for these observations overlapped with those used by VLBI. Given the low declination of the target and hence very low elevations for telescopes in e-MERLIN, the available on-source observing time per observation was limited to ~5 hours and phase-referencing and on-sky amplitude calibration of these data was challenging. To mitigate this, we used a short phase referencing cycle time alternating between 2 min on target and 5.5 min on the phase-reference source J1311−2329, located 26′ from the target. Additional observations every 30 min were made of the nearby ICRF source J1259−2310 . Standard data analysis procedures were undertaken using the NRAO Common Astronomy Software Applications (CASA) packages and utilizing the e-MERLIN data pipeline (*50*). Images of the individual runs provide a noise level between 55 and 90 µJy beam$^{-1}$ and a synthesized beam size of about 180×40 mas. A combined image of all the runs yield a 3-sigma upper limit on the source peak brightness of 60 µJy beam$^{-1}$. When considering only the closest days to the EVN observation, from March 11 to 14, the upper limit is 100 µJy beam$^{-1}$. The nucleus of the host galaxy NGC 4993 is detected on individual epochs with an average flux density and statistical uncertainty of 270±50 µJy. The flux density of NGC 4993 from the combined dataset is 220± 20 µJy.

Estimation of the source size.

   Our image Signal to Noise Ratio (SNR) is limited to around 5, thus determining the size of the radio emitting region is not straightforward. Fitting a circular-Gaussian component in DIFMAP yields a size of 2.9 mas, which is consistent with our naturally weighted un-tapered beam size of 3.5 × 1.5 mas, with major axis position angle of −6 degrees. However, we interpret this as an upper limit because the fitting procedure clearly overestimates the source flux density (93 µJy which is incompatible both with the 60 µJy beam$^{-1}$ upper-limit from our coordinated e-MERLIN observations, and with the value of 47 ± 9 µJy obtained by interpolating published VLA flux densities (*11, 15*) to our observation time and frequency) due to the low SNR. If we fix the total source flux density to values in the range 47 ± 9 µJy, we get best-fitting size measurements in the range 1.3 ± 0.6 mas. As an additional test, we apply various Gaussian tapers to the uv-data to artificially lower our resolution, to verify if we see missed extended flux in the source. While there is some small increase (up to 60 ± 12 µJy beam$^{-1}$ for a 0.5 taper at 20 Mλ, where λ is the wavelength) in the peak brightness, the SNR remains the same. A similar behaviour is seen for the AGN of NGC 4993. Therefore, the increase in peak brightness is a combined effect of the higher noise in the tapered image, and some coherence losses at the longest baselines. Regardless of the uv-taper, the peak brightness in our naturally weighted maps is fully consistent with the total flux density of 47 ± 9 µJy. We interpret this as evidence that the source is at most partially resolved, with a size smaller than or equal to our beam. Within a Bayesian framework, we can estimate the posterior probability on the source size given our peak brightness measurement, taking the information on the source total flux density as a prior. We model the source as an elliptical Gaussian with axes aligned to the East-West and North-South directions. Let $s_x$ and $s_y$ be the full widths at half maxima along such axes. Let $F$ be the total flux, and $F_p$ the measured peak brightness. Applying Bayes' theorem, the posterior probability on the source size and total flux, given the measured peak flux, is given by

$$P(s_x, s_y, F | F_p) = \frac{P(F_p | s_x, s_y, F) P(F) P(s_x, s_y)}{P(F_p)} \quad (S1)$$

Here $P(F_p | s_x, s_y, F)$ is the likelihood of measuring a peak brightness $F_p$ given the sizes $s_x$, $s_y$ and the total flux density $F$, $P(F)$ is the prior on the total flux density, $P(s_x, s_y)$ is the prior on the size and $P(F_p)$ is a normalization constant. We take $P(F)$ to be a normal distribution with mean 47 µJy and sigma 9 µJy. We choose a flat prior on the size, i.e. $P(s_x, s_y) \propto 1$. Since we are actually interested in the marginalized posterior $P(s_x, s_y | F_p)$, we compute the integral of $P(F_p | s_x, s_y, F) P(F)$ over $F$ by Monte Carlo sampling, as follows (see Figure S5 for an illustration of the basic steps). We set up a grid on the two-dimensional parameter space defined by $(s_x, s_y,)$. For each point of the grid, we produce source model images with the given dimension, with total fluxes sampled from the prior $P(F)$. We convolve such images with the beam of our un-tapered, naturally weighted map. We add random realizations of the noise, taken directly from the real image (to account for noise non-gaussianity) to the resulting image, and we record the obtained source peak brightness. Based on such peak brightness samples, we use kernel density estimation at our peak brightness $F_p$ to compute the (un-normalized) integral at each point of the grid. We then normalize the result to obtain the marginalized posterior, which is shown in Figure S3A. To explore the possibility that our measure of the peak brightness is affected by a 10% reduction due to coherence losses at the longest baselines, we repeat the same procedure using a corrected peak brightness $F_p$ = 42/0.9 μJy/beam = 46.7 μJy/beam. The result is shown in Figure S3B. The result of the same procedure, but with a circular Gaussian source model (resulting hence in a 1-dimensional posterior on the source size) is shown in Figure S4. This latter result shows that ~90% of the posterior probability corresponds to sizes smaller than 2.5 mas (in the case without the correction for the 10% loss in coherence). In other words, we can exclude a size larger than 2.5 mas at the 90% confidence level. All results show that the structured jet model is strongly favored with respect to all cocoon models based on these size measurements only, and that the cocoon models with an opening angle larger than 30 deg can be excluded with high confidence.

Structured jet parameter estimation
We fitted a power-law structured jet model (see below) to the 3 GHz, optical and X-ray light curves and to the centroid displacement as measured by comparing our position with those observed with the HSA (*22*). The model parameters are: the core isotropic equivalent kinetic energy $E_c$, energy power law slope $s_1$, core Lorentz factor $\Gamma_c$, Lorentz factor slope $s_2$, core half-opening angle $\theta_c$, interstellar medium (ISM) number density n, electron and magnetic field equipartition parameters $\varepsilon_e$ and $\varepsilon_B$, post-shock electron power law index p and viewing angle $\theta_v$. We fix p = 2.15 as indicated by the measured spectra, and we fix $\varepsilon_e$ to its typical value 0.1. For the fitting, we employ a standard Gaussian likelihood for all data points, with unequal sides in case of asymmetric uncertainties. In case of upper limits, we use a Gaussian penalty with a sigma equal to 10% of the value. The Markov Chain Monte Carlo (MCMC) is performed using the EMCEE Python package (*51*), employing 16 walkers, each running for 30000 steps. We assume uniform priors on $\log(E_c)$, $s_1$, $\log(\Gamma_c)$, $s_2$, $\theta_c$, $\log(\varepsilon_B)$, $\log(n)$ and $\theta_v$. We impose the bounds $\Gamma_c$ < 1000, $\varepsilon_B$ > $10^{-5}$ and $s_1$, $s_2$ < 8 because of model degeneracy beyond these values. Once the MCMC is done, the best fitting values are obtained by performing a principal component decomposition of the posterior samples, measuring the medians and then transforming back to the original base (the light curve of the jet with the resulting parameter values is shown in Figure 3). The uncertainties are computed as the 16th and 84th percentiles of the posterior samples along each axis, thus representing one-sigma confidence ranges. The numerical values are reported in Table S1. Several parameters show degeneracies, as can be seen from the correlations in Figure S1. This means that there are several combinations of parameters compatible with the observed light curves and centroid displacement. The posterior distributions of $\Gamma_c$, $s_2$ and $\log(\varepsilon_B)$ do not show a peak, meaning that any value $\Gamma_c \geq 100$, $s_2 \leq 6$ and $\varepsilon_B \leq 0.03$ corresponds to a possible solution. Some other parameters are rather constrained. The viewing angle is $\theta_v = 15^{+1.5}_{-1.0}$ deg. However, our model does not account for the jet side expansion, which may affect the motion of the observed brightness distribution centroid, therefore the uncertainty on this parameter may be underestimated. Our analysis does not account for the information on the binary inclination that has been derived from the GW analysis. Assuming that the jet is launched in a direction perpendicular to the orbital plane, the binary inclination angle can be related to the jet viewing angle. The binary orbital plane inclination ι derived from the GW analysis, when the information on the host galaxy distance is included, is (*52*) ι = $151^{+15}_{-11}$ deg (90% confidence range). Assuming that the jet axis is perpendicular to the orbital plane, the quantity $\theta_v$ = 180 - ι deg represents the jet axis orientation with respect to the line of sight. Therefore, this range includes our best fitting value 15 deg, even though it is near the edge of the range. Repeating the analysis using the GW results as a prior would yield a better (multi-messenger) estimate of this parameter: we plan to perform such an analysis in a future work (see (*29*) for a fit that includes the GW information).

Comparison of the jet structure with numerical simulations and other results in the literature
Figure S6 shows a comparison of the jet structure determined from our fitting with the results of numerical simulations (*54, 55* their narrow engine case) and with other semi-analytical jet structures from the literature (*12, 13*). All structures feature an approximately constant kinetic energy per unit solid angle within a narrow core, and a steep decrease outside of it. The value of the core kinetic energy density for all the models we consider is consistent within one standard deviation with our result. The core of (*55*), however, is larger than our best-fitting value. Similar considerations apply for the Lorentz factor, with some caveats: given that the shock propagation in the ISM follows a self-similar behaviour, the information on the initial core Lorentz factor is essentially wiped out at the time when the core emission starts to dominate the light curve (i.e. around the peak). As a result, the core Lorentz factor of an off-axis jet can only be constrained to be larger than some value, which in our case is roughly 100. On the other hand, often the Lorentz factor that can be attained in a numerical simulation is limited by the resolution. With these caveats, we consider our Lorentz factor structure to be in agreement with both the numerical simulations and the model from (*13*). In (*12*) the authors do not specify any Lorentz factor angular dependence, presumably because of the said self-similar behaviour, even though the emission of the material moving close to the line of sight does depend on the initial Lorentz factor prior to the deceleration radius.

Choked jet cocoon model parameters
    We model the choked jet cocoon as described below. The light curves can be fitted by several combinations of parameters. The size of the radio image of the cocoon, however, is mainly affected by the half-opening angle (i.e. by its degree of anisotropy), while the other parameters have only a minor impact. We thus choose a single set of parameters, and vary the half-opening angle, compensating the flux difference by changing the electron equipartition parameter $\varepsilon_e$ correspondingly. All models thus have a viewing angle $\theta_v$ = 30 deg, a velocity profile $E(>\Gamma\beta) = E_0(\Gamma\beta)^{-\alpha}$ with $\alpha$ = 6 and $E_0$=1.5 × $10^{52}$ erg, a maximum Lorentz factor $\Gamma_{max}$ = 6 and a minimum ejecta velocity $\beta_{min}$ = 0.89, an ISM number density $n$ = 1.8 × $10^{-4}$ cm$^{-3}$, a post-shock magnetic field equipartition parameter $\varepsilon_B$ = 0.01 and an electron power law slope $p$ = 2.15 (these parameters are similar to those used by (*13*)). The values of the half-opening angle are $\theta_c$ = 30, 45 and 60 deg, and the corresponding electron equipartition parameter values are $\varepsilon_e$ = 0.1, 0.05 and 0.045. The light curves are all the same, so only a single line is plotted in Figure 3.

Comparison of model images with Global VLBI source image
    Since using the CLEAN algorithm on low-SNR data may result in unwanted artifacts in the image, and because our source appears unresolved, we fitted a single 42 μJy point source

component to the data to produce the naturally weighted and untapered map in DIFMAP. We restored the source image by applying the CLEAN algorithm with a "shallow" loop parameter in a large window in order to smooth the noise around the source. The resulting image (Figure 1A) was visually compared to simulated images for the successful jet and choked jet scenarios (Figure 1C-E). To produce these latter images, we computed the predicted surface brightness distribution for the successful jet model and for the three choked jet models described above. These images were read into AIPS, and then convolved with a beam of 3.5 × 1.5 mas with a 0 deg position angle using CONVL. We added noise with rms 8 µJy beam$^{-1}$ (a noise map from the GRB 170817A field, but shifted in position, thus off the target) using COMB. The resulting images (Figure 1) show that the successful jet reproduces the correct peak flux, while the cocoons are resolved and thus lead to a lower measured peak flux. Cocoons with $\theta_c$ = 45 and 60 deg have essentially identical appearance. Only the cocoon with $\theta_c$ = 30 deg can still be regarded as marginally consistent with the data, its peak brightness being ~ 32 µJy beam$^{-1}$ (with that particular noise realization). However, the upper limit on the size excludes this model with 75% confidence or better (Figures S3 and S4).

Jet dynamics and parametrization of its angular structure

In order to compute the expected light curve and radio image of the jet, we need to know the dynamics of the jet expansion and deceleration into the interstellar medium (ISM). We assume, for simplicity, that the jet structure is axisymmetric, i.e. it depends only on the polar angle $\theta$ (i.e. on the angular distance from the jet axis) and not on the azimuthal angle $\varphi$. Let us consider a "jet element", a small portion of jet material between $\theta$ and $\theta + d\theta$. As long as the jet element moves faster than the local sound speed, it is out of causal contact with the rest of the jet, and thus its dynamics in that phase must depend only on its initial bulk Lorentz factor $\Gamma_0(\theta)$, its kinetic energy per unit solid angle $dE(\theta)/d\Omega = dE(\theta)/2\pi \sin\theta d\theta$ and on the number density $n$ of the ISM. Later on, as soon as the speed of sound waves close to the jet element becomes comparable to that of the expansion, the dynamics should start being affected by the surrounding jet elements, and thus should depend on the global jet structure, which makes it a much more difficult problem if analytical methods are to be employed. Nevertheless, numerical simulations indicate (*53*) that the energy transport in this phase is rather slow, so that the dynamics does not change drastically until at relatively late time. For these reasons, we neglect the effect of causal contact between neighbouring jet elements during the expansion, and we assume that each jet element expands and decelerates as if it were part of a spherical explosion with an isotropic equivalent kinetic energy $E_{K,iso}(\theta) = 4\pi \, dE(\theta)/d\Omega$ expanding adiabatically into the ISM. This assumption is common to most published models of the structured jet dynamics used in the modelling of GRB 170817A so far (*11, 12, 54, 55*). Under these assumptions, the dynamics of each jet element can be computed by requiring conservation of energy as it sweeps the ISM (*56*). We parametrize the jet structure defining two functions:

$$E_{k,iso}(\theta) = \frac{E_c}{1 + (\theta/\theta_c)^{s_1}}$$

$$\Gamma(\theta) = 1 + \frac{\Gamma_c - 1}{1 + (\theta/\theta_c)^{s_2}}$$
(S2)

which describe respectively the isotropic equivalent and initial Lorentz factor (i.e. the initial speed) as functions of the polar angle $\theta$. This parametrization implies that these two quantities are constant within a jet core of half-opening angle $\theta_c$ and they decrease as power laws (of index $s_1$ and $s_2$, respectively) as a function of $\theta$ outside of it. The quantities $\theta_c$, $E_c$, $\Gamma_c$, $s_1$ and $s_2$ are the parameters that define the structure.

Isotropic outflow dynamics and parametrization of its radial structure

The dynamics of the isotropic outflow expansion and deceleration are different from those of a jet element, because the outflow is assumed to have a radial velocity structure, i.e. the ejecta in the outflow do not move all at the same velocity, but instead the kinetic energy is radially distributed between ejecta with a range of velocities. The outer, fastest ejecta shell first drives a shock in the ISM, which is then progressively reached by the inner, slower shells, which gradually contribute their energy to the shocked region. This effective, progressive energy injection results in a slower deceleration with respect to that of a shell with a single velocity. As soon as the slowest material has entered the shocked region, the deceleration turns back to follow the same laws as that of a uniform shell, because the energy injection from behind has stopped. The analytical solution for the dynamics can be derived again based on energy conservation (*57*).

The radial velocity profile can be specified by quantifying the amount of kinetic energy that is contained in the ejecta faster than each given velocity $v$, which we indicate as $E(>v)$. This must be a decreasing function of $v$ (by definition), and the faster it decreases, the smaller the fraction of energy contained in the fast tail of the ejecta. We employ the same parametrization as (*13*), namely

$$E(v) = E_0 \times \begin{cases} 0 & v/c > \beta_{max} \\ \left(\dfrac{v/c}{\sqrt{1-v^2/c^2}}\right)^{-\alpha} & \beta_{min} < v/c < \beta_{max} \\ \left(\dfrac{\beta_{min}}{\sqrt{1-\beta_{min}^2}}\right)^{-\alpha} & v/c < \beta_{min} \end{cases} \qquad (S3)$$

which means that no significant energy is contained in ejecta faster than $\beta_{max}c$ or slower than $\beta_{min}c$, where $c$ is the speed of light. The parameter $\alpha$ controls how steep is the decrease of the energy with velocity, the parameter $E_0$ equals the total energy multiplied by $\left(\beta_{min}/\sqrt{1-\beta_{min}^2}\right)^{\alpha}$. We also introduce an additional parameter, namely the outflow half-opening angle $\theta_c$. This allows us to account for the possible anisotropy in the cocoon energy distribution.

We assume that both the jet and the isotropic outflow material emit synchrotron radiation and we neglect the contribution of Compton scattering of synchrotron or thermal photons, which is not expected to contribute in the considered bands. Since the data do not contain any sign of synchrotron self-absorption in the source (*11*), we consider it to be optically thin, avoiding the computation of the effects of self-absorption for simplicity. In order to compute the image, we need to specify the shock profile, i.e. how the energy and number density of electrons and magnetic field are distributed in the volume behind the shock. For the jet, in accordance with our assumptions on the dynamics, we assume that the shock profile at any fixed time, at any fixed polar angle is described by the Blandford-McKee self-similar spherical impulsive blast-wave solution (*58*) with energy equal to the isotropic equivalent energy at that angle. For the spherical outflow, as long as the reverse shock is continuously crossed by the slower material, the shock profile does not resemble that of an impulsive explosion: the emission rather comes from a thin region comprised between the forward shock (which propagates into the ISM) and the reverse shock (which propagates back into the ejecta). As soon as the slowest ejecta have crossed the reverse shock, the profile turns back to that of an impulsive explosion. Since our observation took place after the peak of the light curve (i.e. after all the ejecta have crossed the reverse shock), we model the shock profile with the Blandford-McKee impulsive solution in this case as well.

Computation of the model images

Given all the assumptions described in the preceding sections, we compute the comoving synchrotron emissivity $j_{\nu'}$ on a 3D grid containing the emitting volume, at times (measured in the ISM frame) that correspond to the light cone of an observer on Earth who receives the photons at time $t_{obs}$ = 207.4 d after GW170817, i.e. at the time of our global-VLBI observation. The emissivity is computed for emission at the comoving frequency corresponding to $\nu_{obs}$ = 5 GHz in the observer frame (taking into account both the Doppler shift and the cosmological redshift), and it is multiplied by the square of the Doppler factor to account for the relativistic transformation to the observer frame. Integration of the emissivities along the line of sight then results in a map of surface brightness, which is the source image (Figure 2).

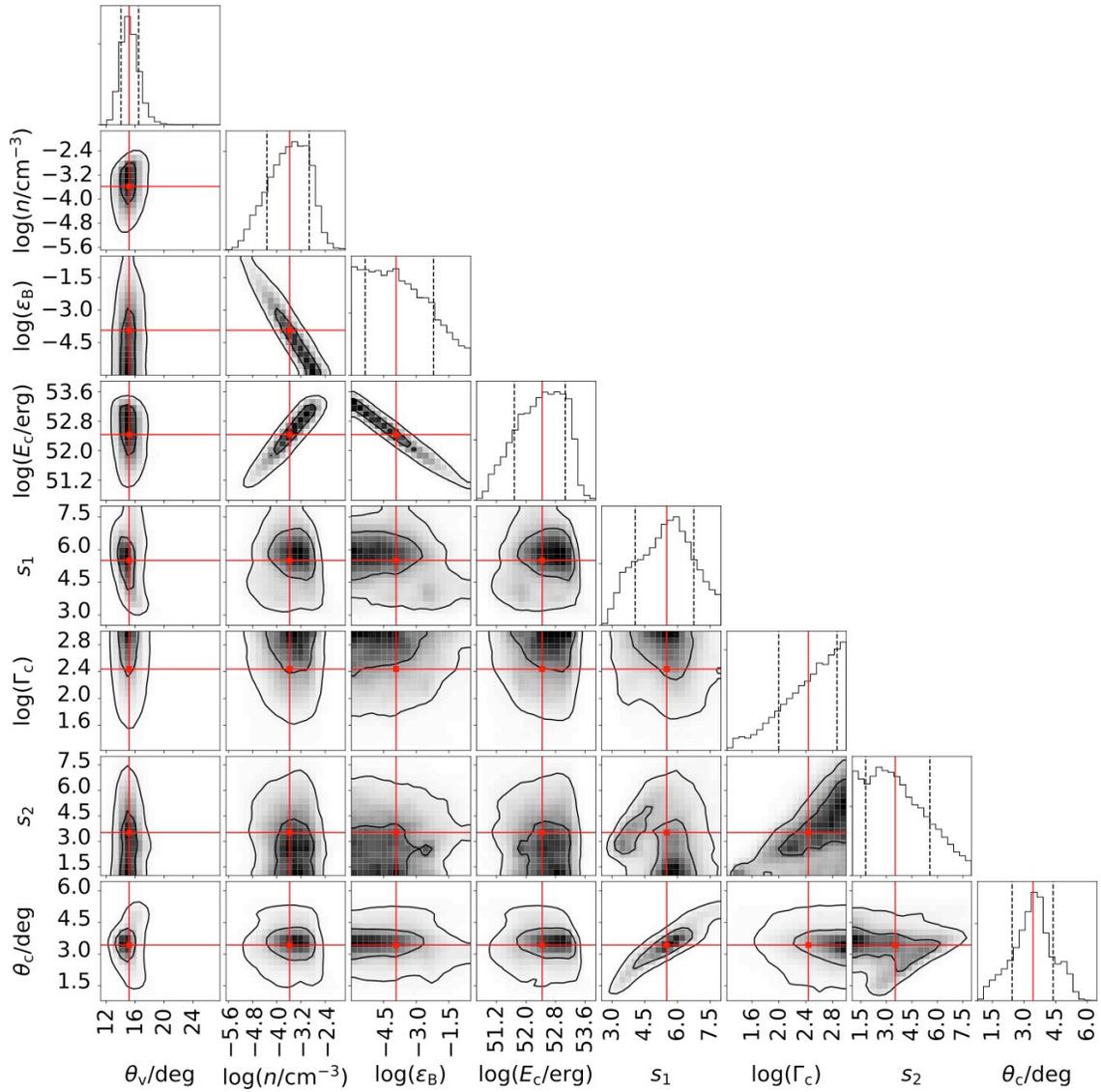

**Fig. S1. Corner plot showing the results of our Markov Chain Monte Carlo parameter estimation for the structured jet model**. The histograms on the diagonal show the marginalised posterior densities for each parameter (vertical dashed lines represent the 16$^{th}$ and 84$^{th}$ percentiles). The remaining plots show the 2D joint posterior densities of all couples of parameters, with 1$\sigma$ and 2$\sigma$ contours shown by black solid lines, and our best fitting parameters shown by red squares and lines. Values are listed in Table S1.

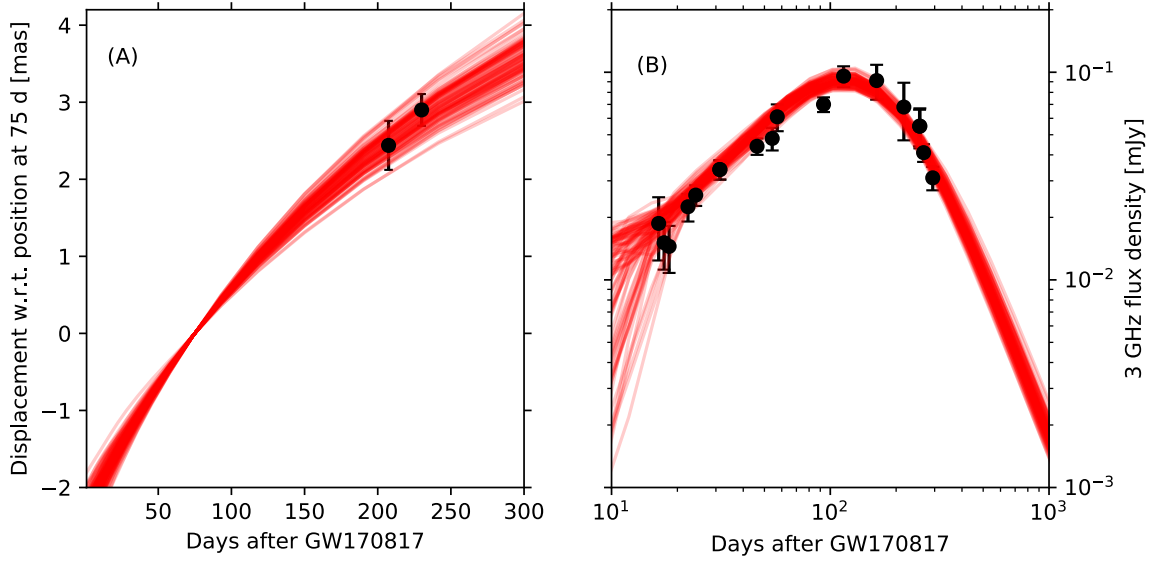

**Fig. S2. Centroid displacement and 3 GHz light curve of 100 posterior samples from our Markov Chain Monte Carlo parameter estimation for the structured jet model**. Panel (A): projected image centroid displacement in mas, as a function of the time since GW17817. The black points with error bars show the displacement measured comparing our position with those by the HSA *(22)*. The red solid curves represent the centroid displacement with time corresponding to 100 randomly selected posterior samples. Panel (B): the same as for panel (A), but showing the 3 GHz radio light curve.

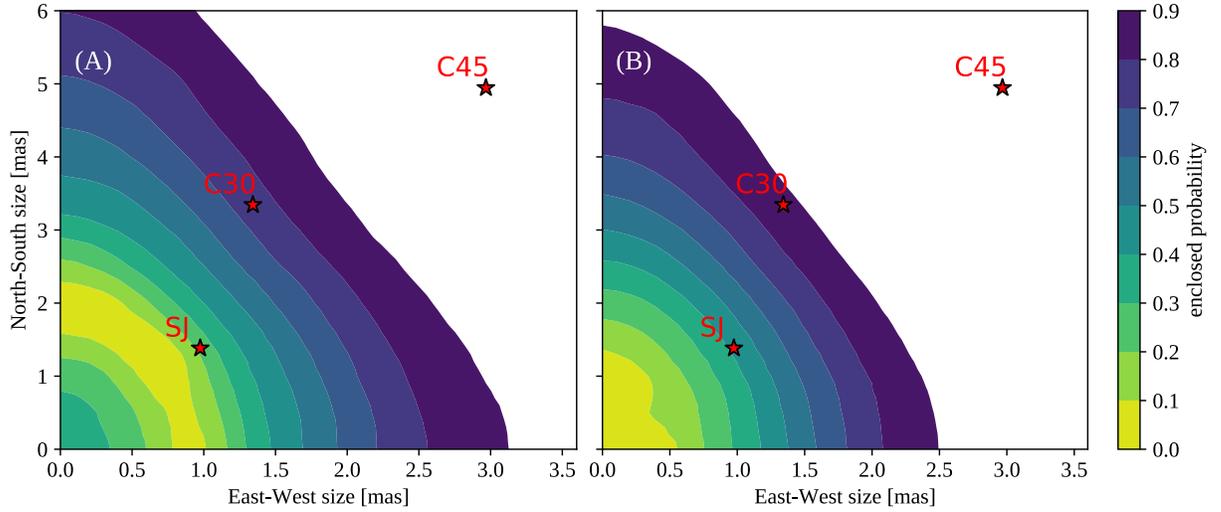

**Fig. S3. Source size constraint.** The filled contours show the posterior probability on the size of the source, modeled as an elliptical Gaussian with axes aligned to the north-south and east-west directions in the plane of the sky in our observations. Each color transition represents a 10% increase in the enclosed posterior probability. The peak is represented by the lightest area. The red stars represent the sizes of elliptical Gaussian fits to three of our models (SJ is the successful jet, C30 the cocoon with $\theta_c = 30$ deg, and C45 the cocoon with $\theta_c = 45$ deg). Panel (A): no coherence loss correction; panel (B): measured peak flux increased to account for a 10% brightness underestimate due to coherence losses at the longest baselines.

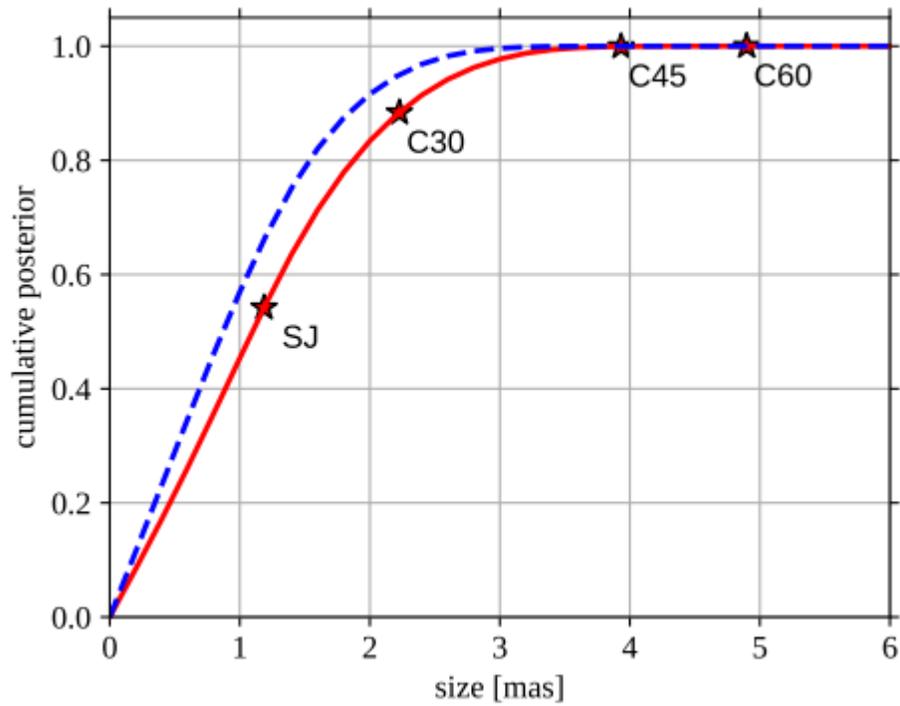

**Figure S4: Equal-axial-ratio source size constraint.** The red line represents the cumulative posterior probability on the source size, assuming a circular Gaussian model. The blue dashed line is the same, but accounting for a 10% peak brightness loss in our measurement due to coherence losses at the longest baselines. The red stars represent the model sizes (same labels as Figure S3; here we also show C60 which represents the cocoon with $\theta_c$ = 60 deg) obtained by fitting a circular Gaussian to each model image.

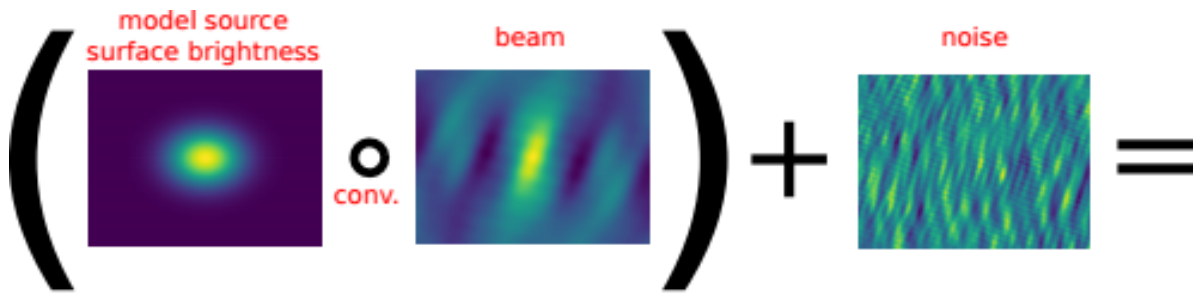

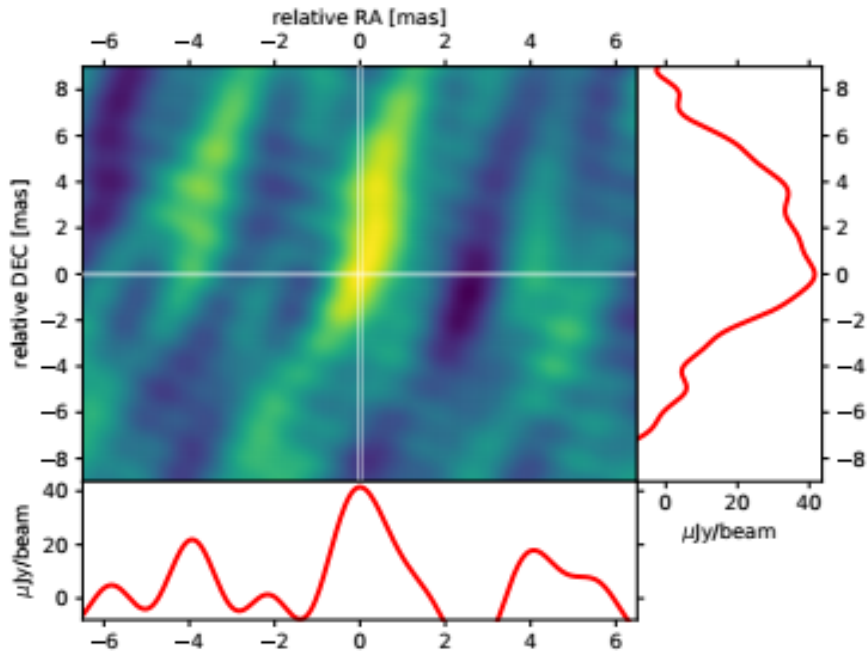

**Figure S5: Depiction of our Monte Carlo posterior probability calculation**. To produce the needed peak brightness samples, the model image is convolved to the beam, and a noise realization is added to obtain a simulated source image (an example of which is shown above, along with plots of the brightness distribution along the central axes, highlighted in white), whose peak brightness is recorded. The procedure is repeated for a large number of noise realizations (obtained by placing the convolved source image at random position of the noise image) and for a large number of source total flux densities sampled from the prior. Gaussian kernel density estimation is then used on the produced peak brightness samples to compute the un-normalized, marginalized posterior for each size. Varying the source image size on a grid we obtain the posterior probabilities shown in Figures S3 and S4.

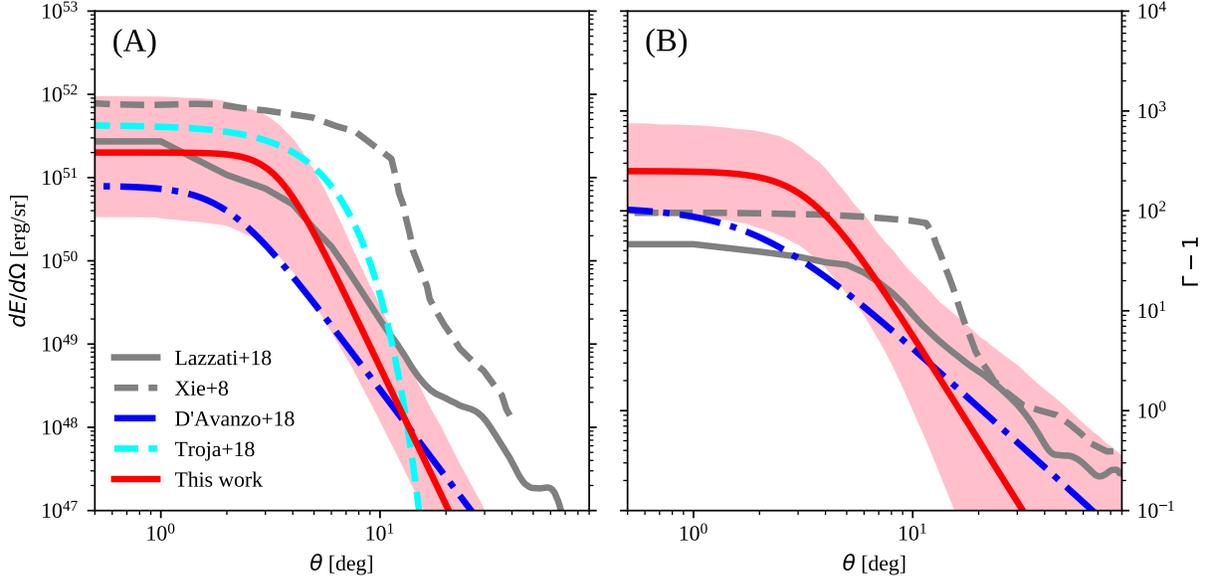

**Figure S6:** Comparison of the jet structure we derive with those from numerical simulations (thick solid and thick dashed grey lines, respectively from (*54*) and (*55*), as labeled) and with other analytical structures from the literature (dashed cyan and dot—dashed blue line for (*7*) and (*13*) respectively). Our best fitting models are shown by the red solid lines, with the pink shaded region representing the one-sigma uncertainty (i.e. values between the 16$^{th}$ and 84$^{th}$ percentile of those span by structures obtained from our MCMC posterior samples). The cyan dashed and blue dot-dashed lines represent the jet structures from (*12*) and (*13*) respectively, as labeled. (A) shows the jet kinetic energy per unit solid angle, while (B) shows the initial fluid Lorentz factor. No initial Lorentz factor is specified in the model of (*12*).

**Table S1.** Results of our parameter estimation for the structured jet model.

| Parameter | Best fitting value | One sigma range |
|---|---|---|
| Log($E_c$/erg) | 52.4 | (51.7, 53.0) |
| $s_1$ | 5.5 | (4.1, 6.8) |
| Log($\Gamma_c$) | 2.4 | (2.0, 2.9) |
| $s_2$ | 3.5 | (1.8, 5.6) |
| $\theta_c$/deg | 3.4 | (2.4, 4.4) |
| Log($\varepsilon_B$) | -3.9 | (-5.4, -2.2) |
| Log($n$/cm$^{-3}$) | -3.6 | (-4.3, -2.9) |
| $\theta_v$/deg | 15 | (14, 16.5) |

**Table S2**: J2000 measured positions of GRB 170817A, the NGC 4993 AGN and the reference sources J1312-2350 and J1311-2329, both before (second and third columns) and after (fourth and fifth columns) shifting the reference system to match the position of J1312-2350 with its known ICRF coordinates (*59*). The statistical uncertainty on the position of J1312-2350 has been added in quadrature to all other position uncertainties after the shift. The systematic error on the position of GRB 170817A (reported in parentheses) due to atmospheric fluctuations can be estimated, based on simulations (*60*), to be approximately 0.035 mas in RA and 0.14 mas in Dec. An additional source of systematic error is the uncertainty on the absolute position of J1312-2350, which is of the order of 1 mas (as reported by the AstroGeo Center, see *61*). The latter should cancel out, though, in the computation of the displacement of our source with respect to the positions measured with the HSA (*22*).

| | With our original reference source J1311-2329 (statistical only) | | After applying a (1.1625 mas, -0.905 mas) shift to match J1312-2350 position with its ICRF coordinates (*59*) | |
|---|---|---|---|---|
| **Source Name** | **RA** | **Dec** | **RA** | **Dec** |
| **J1311-2329** | $13^h11^m37^s.4139866$ (assumed) | $-23°29'56".646509$ (assumed) | $13^h11^m37^s.4140641 \pm 0^s.0000014$ | $-23°29'56".645604 \pm 0".000037$ |
| **J1312-2350** | $13^h12^m48^s.7580225 \pm 0^s.0000014$ | $-23°50'46".953905 \pm 0".000037$ | $13^h12^m48^s.7581000 \pm 0^s.0000014$ | $-23°50'46".953000 \pm 0".000037$ |
| **NGC 4993 AGN** | $13^h09^m47^s.6939463 \pm 0^s.0000031$ | $-23°23'02".32091 \pm 0".00018$ | $13^h09^m47^s.6940238 \pm 0^s.0000034$ | $-23°23'02".32000 \pm 0".00018$ |
| **GRB 170817A** | $13^h09^m48^s.0687231 \pm 0^s.000014$ | $-23°22'53".39167 \pm 0".00025$ | $13^h09^m48^s.06880 \pm 0^s.00002$ ($\pm 0^s.000002$) | $-23°22'53".390765 \pm 0".00025$ ($\pm 0".00014$) |

16